\RequirePackage{fix-cm}
\documentclass[smallextended]{svjour3}       % onecolumn (second format)
\smartqed  % flush right qed marks, e.g. at end of proof
\usepackage{graphicx}
\def\ph2{{\it p}-H$_2$}
\def\od2{{\it o}-D$_2$}
\begin{document}
\title{Quasi-2D  H$_2$: On the verge of turning superfluid?}
\author{Massimo Boninsegni}
\institute{
              Department of Physics, University of Alberta, Edmonton, Alberta, Canada, T6G 2E1\\
              \email{m.boninsegni@ualberta.ca}
}
\date{Received: date / Accepted: date}

\maketitle

\begin{abstract}
First principle computer simulations of a thin parahydrogen film adsorbed on a silica substrate  at low temperature (below 6 K) yield no evidence that the top layer is 
liquid and/or in the proximity of a superfluid transition, as claimed in recent experimental work [T. Makiuchi {\em et al.}, Phys. Rev. Lett. {\bf 123}, 245301 (2019)]. Computed values of first and second layer completion densities are in quantitative agreement with experiment, but as observed also on other substrates, the top layer is an insulating crystal, quantum-mechanical exchanges of molecules are non-existent, and the overall physical behavior of the system can be understood largely along classical lines.

\end{abstract}

\section{Introduction}
\label{intro}
The search for a naturally occurring condensed matter system other than helium, capable of displaying the stunning phenomenon of superfluidity, has motivated decades of experimental and theoretical investigation of condensed parahydrogen (\ph2). One naturally thinks of it  as a potential second superfluid, for its elementary constituents, namely \ph2 molecules, are composite bosons of spin $S=0$ with a  mass equal to one half of that of a $^4$He atom.\\ Considering  fluid \ph2 as a non-interacting gas, Ginzburg and Sobyanin \cite{gs} proposed that Bose-Einstein Condensation (BEC) ought to occur at a temperature $T\sim 6$ K. Such a simple model yields an equivalent temperature for $^4$He $\sim$ 3 K, remarkably close to that at which BEC is observed experimentally, with the concurrent onset of superfluidity; it seems thus plausible that the same physical behavior might occur in \ph2.
\\ \indent
Experimental investigation spanning a few decades \cite{bretz,maris,rail,schindler,sokol,chan}, however, has failed to observe the putative superfluid (SF) phase of \ph2, which, unlike $^4$He, solidifies at temperature $T=13.8$ K, i.e., well above that at which the superfluid transition should take place.
\\ \indent 
{\em No controversy} exists at the theoretical level about the fact that the different behavior of \ph2 and $^4$He is a direct consequence of the very different relative importance of interparticle interactions, imparting to \ph2 a strong propensity to crystallize \cite{boninsegni04,boninsegni13}. Molecular localization, and the consequent absence of superfluidity, is predicted even in confinement \cite{omiyinka,screw}, disorder \cite{turnbull}, or in thin films intercalated with a regular array of impurities \cite{boninsegni05,boninsegni16}. It seems fair to state that no credible scenario of {\em bulk} \ph2 superfluidity is presently being investigated, or even discussed; the only quantitative prediction of SF behavior of \ph2 has been made for small clusters (few tens of \ph2 molecules), at temperatures of the order of 1 K \cite{sindzingre,kwon,fabio,fabio2,boninsegni20}; some experimental evidence of SF behavior of these clusters has actually been obtained \cite{grebenev,li,raston}. 
\\ \indent
One might wonder ``how far'', so to speak, bulk  \ph2 is from turning SF; is it ``on the verge'' of undergoing such a transition, one clever idea or experimental trick away from circumventing solidification, or is it intrinsically prevented from displaying superfluidity, due to a particular combination of particle mass and interaction?  
There are strong indications in favor of the latter scenario, one that no amount of effort or ingenuity on the part of experimenters may overcome. Besides the above-mentioned experimental failure to measure a SF response, in very different experimental conditions, no credible theoretical evidence of a possible bulk metastable (super)fluid  phase has been reported so far \cite{boninsegni04,boninsegni18}.  Solidification occurs in \ph2 as a direct consequence of the suppression of quantum-mechanical exchanges, which are known to play a crucial role in the crystallization of Lennard-Jones-like Bose system \cite{role}. Exchanges are hindered in \ph2 by  the relatively large value of the hard core diameter of the intermolecular interaction \cite{boninsegni18}, {\em not} by the depth of its attractive well (a long held, erroneous belief).
\\ \indent
In a recent experimental paper, however, the assertion has been made that the top layer of a thin (few layers thick) \ph2 film adsorbed on a glass substrate is, in fact, in the near proximity of a phase transition to a SF phase \cite{ikr}. This conclusion is based on measurements  of the elastic response of films of \ph2 (as  well as HD and D$_2$) adsorbed inside gelsil -- a porous glass which can be regarded as a network of interconnected cylindrical channels\footnote{Theoretical evidence suggests that the properties of an adsorbed layer of \ph2 on the inner surface of a cylinder of such a large diameter, are essentially identical with those on a flat substrate. See, for instance, Ref. \cite{omiyinka}.} of average diameter $\sim 40$ \AA. Specific anomalies in the observed elastic response are interpreted in Ref. \cite{ikr} as signaling the onset of different physical regimes of diffusion of surface molecules, culminating with their freezing into a localized state at a temperature $T\sim 1$ K, whereupon the layer crystallizes.
The implication is that the top layer of the film may remain in a liquid-like phase, in which molecules experience a rather high mobility, down to a temperature fairly close to that at which a Bereszinskii-Kosterlitz-Thouless SF transition  ought to occur, based on the well-known universal jump condition \cite{nk}, assuming a two-dimensional (2D) density equal to the equilibrium value \cite{boninsegni04}. 
\\ \indent
The scenario laid out in Ref. \cite{ikr}, while certainly intriguing, is at variance with first principle theoretical studies of \ph2 films on various substrates, including weakly attractive ones, showing that adsorption takes place through completion of successive {\em solid } adlayers, whose melting temperature is close to 7 K \cite{melting}, i.e., significantly higher than what implied  in Ref. \cite{ikr}. Those studies have yielded no indication of any liquid-like behavior of the top layer. \\ \indent
In order to provide a theoretical check of the predictions made in Ref. \cite{ikr}, as well as to gain additional theoretical insight, we have carried out first principle computer simulations at low temperature (down to $T$=0.5 K) of a thin (up to two layers) \ph2 film adsorbed on a glass substrate. Our simulations are based on standard Quantum Monte Carlo (QMC) techniques. 
We made use of an accepted pair potential to describe the interaction between \ph2 molecules, while for the interaction between a \ph2 molecule and the substrate we utilized  the simple ``3-9'' potential, with coefficients adjusted by starting from the commonly adopted values for helium, and modifying them to describe \ph2, using the Lorentz-Berthelot combining rules, as done in previous work \cite{omiyinka}. \\ Despite its relative crudeness, this microscopic model turns out to be quite effective in reproducing important experimental observations, such as values of coverage at which the first and second adsorbed layers are completed. However, our results lend no support to the contention of Ref. \cite{ikr}. Rather, we arrive at the same conclusions of  all similar studies of condensed \ph2, which have shaped our current understanding of the system. 
\\ \indent
Quantum-mechanical exchanges, both among molecules in different layers as well as in the same layer, are all but non-existent, all the way down to the lowest temperature considered here. This makes the entire contention that the top adlayer may be ``on the verge'' of turning superfluid,  downright untenable. Rather, we find both a monolayer as well as the second layer to be in an insulating crystalline phase, at temperatures as high as 6 K. Structure and energetics of the adsorbed film display little or no change as the temperature is lowered from 6  to 0.5 K. In summary, our simulations, based on the currently accepted microscopic model of condensed \ph2, one that accurately accounts for a great deal of observed properties of its bulk phase (see, for instance, Ref. \cite{dusseault}), are at variance with the interpretation proposed in Ref. \cite{ikr} of the elastic anomalies observed therein. More generally, they reaffirm the notion that the study of adsorbed \ph2 films is scarcely a promising avenue to the observation of superfluidity.
\\ \indent
The remainder of this manuscript is organized as follows: in sec. \ref{model} we describe the microscopic model adopted in this study and  offer a brief description of the computational methodology adopted; we illustrate our results in sec. \ref{else}, and outline our conclusions in sec. \ref{concl}.

\section{Model and methodology}\label{model}
We consider an ensemble of $N$ \ph2 molecules, regarded as point-like spin-zero bosons, moving in the presence of a smooth, flat  substrate. In the coverages and temperature ranges considered here, up two \ph2 layers form. 
The system is enclosed in a simulation cell shaped as a  cuboid,  with periodic boundary conditions in all directions (but the length of the cell in the $z$ direction can be considered infinite for all practical purposes). The flat (glass) substrate occupies the $z=0$ face of the cuboid, whose area is $A$. The nominal coverage $\theta$ is given by $N/A$.\\ 
The quantum-mechanical many-body Hamiltonian reads as follows:
\begin{eqnarray}\label{u}
\hat H = -\sum_{i}\lambda\nabla^2_{i}+\sum_{i}U({z}_{i})+\sum_{i<j}v(r_{ij}).
\end{eqnarray}
where ${\bf r}_i\equiv (x_i,y_i,z_i)$ is the position of the $i$th molecule, $r_{ij}\equiv|{\bf r}_i-{\bf r}_j|$ and $\lambda=12.031$ K\AA$^{2}$.
The first and second sums run over all the $N$ \ph2 molecules, while the third runs over all pairs of molecules. The function $v(r)$ is the accepted Silvera-Goldman \cite{SG} potential, which describes the interaction between two \ph2 molecules, whereas 
$U$ describes the interaction of a \ph2 molecule with the glass substrate. We use here the so-called ``3-9'' potential:
\begin{equation}\label{39}
U({\bf r})\equiv U(z) = \frac{D}{2}\ \biggl \{ \biggl (\frac{a}{z}\biggr )^9 -3 \biggl (\frac{a}{z}\biggr )^3\biggr\}
\end{equation}
where $D$ is the depth of the attractive well experienced by a molecule in the vicinity of the substrate, while $a$ can be regarded as the classical equilibrium distance of a molecule from the substrate, a strong repulsion intervening at shorter distances. The parameters $D$ and $a$ are adjusted to reproduce as closely as it is allowed by such a simple expression, the interaction of an atomic or molecular species with a given substrate.  To our knowledge, there are no accepted values for these parameters for the case of \ph2 and silica; on the other hand, there have been numerous studies of helium adsorption on such a  substrate, and there seems to be consensus that the value of $D$ for that case \cite{pricaupenko,krotscheck} should be $\sim 100$ K, while $a\sim 2.2$ \AA \ \cite{vidali}. Because the potential form (\ref{39}) is the result of  the integration of a Lennard-Jones potential over a semi-infinite slab, we can use  Lorentz-Berthelot combining rules to obtain the corresponding parameters for the \ph2-silica interaction, namely $D=232$ K and $a=2.37$ \AA, which we have used in all of the simulations carried out in this work.
\\ \indent
As mentioned above, the substrate itself is considered smooth and flat, i.e., its  corrugation is neglected.
In reality, of course, the substrate is irregular, causing the localization of molecules at random locations, and the ensuing appearance of a stable, low-coverage disordered insulating phase, which is obviously lost in a  simulation based on a flat substrate model. As the coverage is increased, however, a first-order transition to a regular (monolayer) crystal is expected, similar to the commensurate-incommensurate transition on regular, corrugated substrates \cite{crespi,nho1}. As further layers are adsorbed, the corrugation of the substrate should become less relevant, i.e., a flat substrate model should capture the essential physics of the problem, the top layer mostly experiencing the corrugation of the lower layer.
\\ \indent
We carried out QMC simulations of the system described in section \ref{model} using the worm algorithm in the continuous-space path integral representation, specifically a canonical implementation of the algorithm in which the total number of particles $N$ is held constant, in order to simulate the system at fixed coverage \cite{fabio,fabio2}.
We obtained results in the temperature interval $0.5$ K $ \le T \le 6$ K, in a range of \ph2  coverage 0.057 \AA$^{-2}$ $\le \theta \le 0.167$ \AA$^{-2}$.
Details of the simulation are standard; we made use of the fourth-order approximation for the high-temperature density matrix (see, for instance, Ref. \cite{jltp}), and all of the results quoted here are extrapolated to the limit of time step $\tau\to 0$. In general, we found that a value of the time step equal to $3.1\times 10^{-3}$ K$^{-1}$ yields estimates indistinguishable from the extrapolated ones, within statistical uncertainties. We carried out most of the calculations whose results are shown here with a total number of particles equal to either 150 or 340, the latter number mainly utilized for calculations for a two-layer film  of coverage $\theta=0.155$ \AA$^{-2}$. We estimate finite-size corrections to the energy to amount to less than 0.1\%.
\\ 
\indent 
In principle, of course, \ph2 molecules  are identical, and therefore, in the presence of more than one adsorbed layer, no conceptual distinction can be drawn between molecules  in different layers. However, in all of the simulations that we have carried out of a 2-layer system we have consistently  observed  that inter-layer hopping and/or quantum-mechanical exchanges of molecules in different layers, are {\em exceedingly infrequent}. It is therefore an  excellent approximation to regard the two layers as two distinct ``species'', which  allows us to compute separately their energetic contributions, as well as focus on the possible superfluid response of the top layer,  of special interest here. We have made this approximation in most the calculations aimed at studying the SF response of the top layer, whose results are presented here. 
\section{Results}\label{else}
\begin{figure}[h]
\centering
\includegraphics[width=\linewidth]{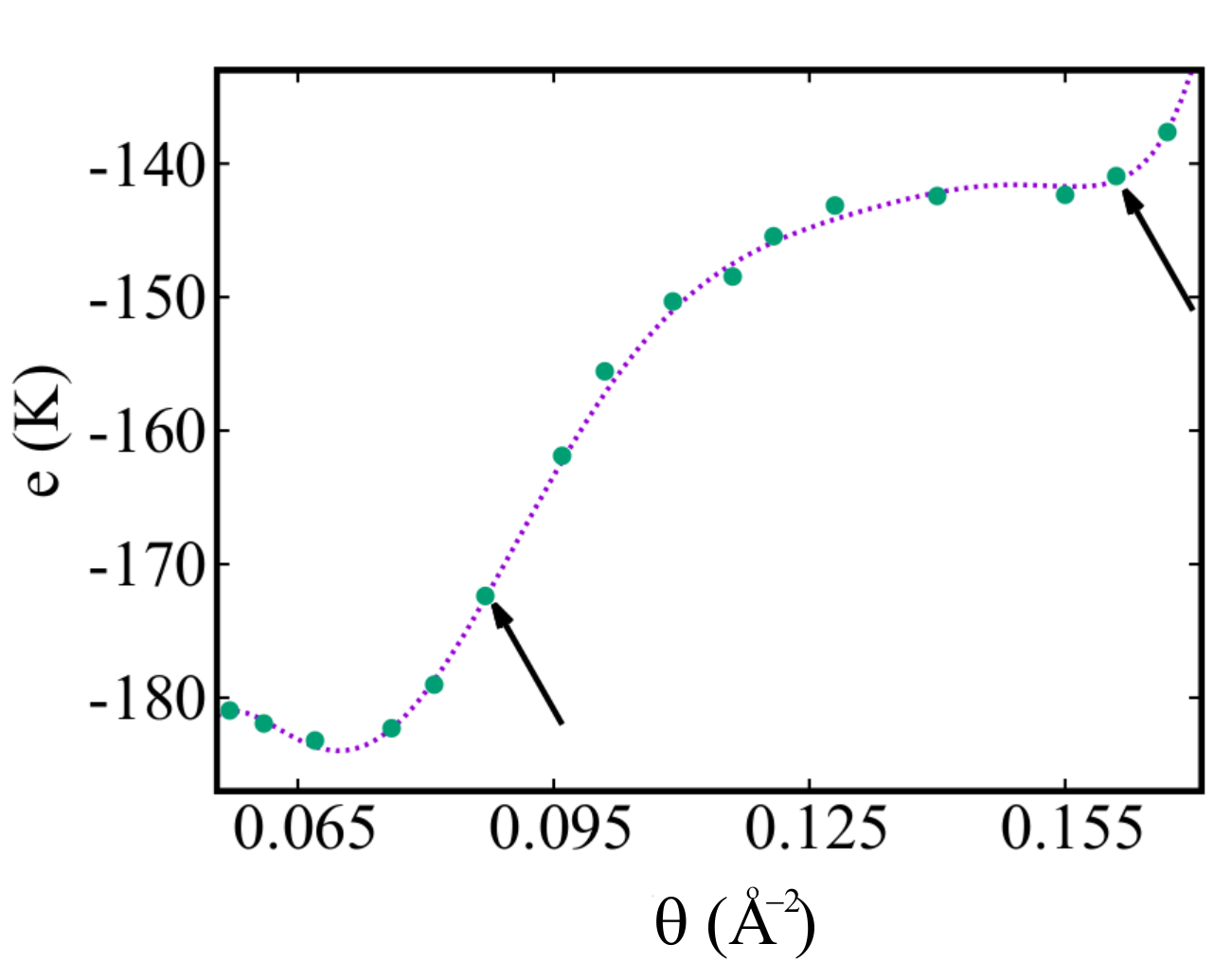}
\caption{Energy per \ph2 molecules (in K) as a function of coverage (in \AA$^{-2}$), at temperature $T=1$ K. 
Dotted line is obtained as a piece-wise cubic spline  fit  to the data. Arrows point to changes of inflection of the curve. Statistical errors are smaller than symbol sizes.}
\label{f1}
\end{figure}

As mentioned in Sec. \ref{model}, we make use of a very simple model for the interaction between a \ph2  molecule and the glass substrate, with parameters whose values are adjusted fairly crudely. Therefore, the first important test of the model that we utilize consists of checking whether it is sufficiently reliable, i.e., if it quantitatively accounts for the most important experimental observations. 
\\ \indent 
Fig. \ref{f1} shows the computed energy per \ph2 molecule at temperature $T=1$ K. The dotted line through the data is a guide to the eye; it is a piece-wise cubic spline fit to the data, and its only purpose is that of helping to identify the points of inflexion. These results can be regarded as essentially ground state estimates, as no significant dependence on the temperature is detectable below $T\sim 4$ K, within our statistical uncertainties. This is very similar to what observed in analogous studies of \ph2 adsorption on other substrates, e.g., lithium \cite{melting} or graphene \cite{dusseault2}.

\begin{figure}[h]
\centering
\includegraphics[width=\linewidth]{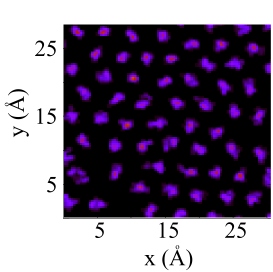}
\caption{Density snapshot of the top layer of a two-layer system of total coverage $\theta=0.155$ \AA$^{-2}$ at temperature $T=6$ K.The density of the bottom layer is 0.088 \AA$^{-2}$. The simulated system comprises altogether $N=150$ particles.}
\label{f2}
\end{figure}

\begin{figure}[h]
\centering
\includegraphics[width=\linewidth]{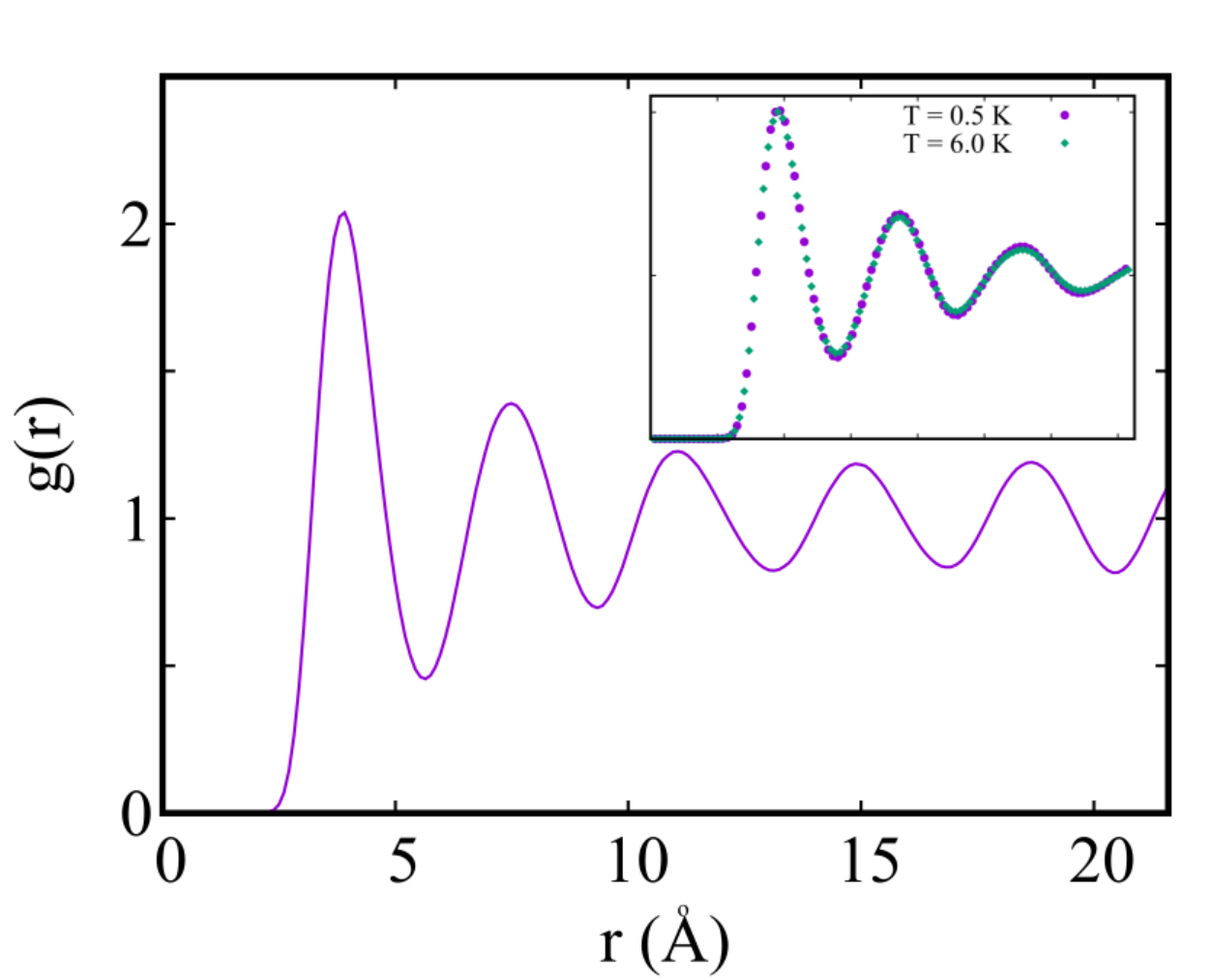}
\caption{ Reduced pair correlation function (i.e., integrated over the direction perpendicular to the substrate) for the top layer of a two-layer \ph2 film. The overall coverage is $\theta=0.155$ \AA$^{-2}$, while the density of the bottom layer is 0.088 \AA$^{-2}$. Inset shows a comparison of the results at the two different temperatures $T=0.5$ and $6$ K.}
\label{f3}
\end{figure}

The curve  $e(\theta)$ features an absolute minimum at a coverage $\theta_e\approx 0.070$ \AA$^{-2}$, only slightly above the purely 2D value \cite{boninsegni04} of 0.067 \AA$^{-2}$. This corresponds to the lowest coverage for which a stable monolayer film can form on the smooth substrate considered here. The fact that the equilibrium density is slightly above that of purely 2D \ph2 is attributable to the zero-point motion of the molecules in the direction perpendicular to the plane, which acts to soften the short-range repulsion of the inter-molecular potential at short distance. The monolayer film can be compressed up to a maximum of $\sim 0.087$ \AA$^{-2}$, at which point the curvature changes (lower arrow in Fig. \ref{f1}), signaling the onset of  second layer promotion, with the formation of clusters of molecules on top of the first layer.
As $\theta$ is increased, the curve retains its negative curvature up to $\theta\sim 0.157$ \AA$^{-2}$, at which point there is another point of inflexion, the curvature turns positive again and a compressible second layer forms. The question is  how consistent all of this is with experimental observation.
\\ \indent
According to Ref. \cite{ikr}, the coverage at which a monolayer is formed is estimated to be $n_1=14.5\ \mu$moles/m$^2$ = 0.087 \AA$^{-2}$, whereas the second layer forms at $n_2\approx 1.8 n_1 = 0.157$ \AA$^{-2}$. Our results seem therefore in remarkable agreement with experiment, if we consider that our computed equilibrium density pertains to a flat substrate. The actual silica is expected to be irregular, and therefore, as explained above, one may expect one or more low coverage, disordered insulating phases, reflecting the random topography of the substrate, with a first-order phase transition to an incommensurate layer as the coverage is increased. This is qualitatively similar to what occurs on a substrate such as graphene \cite{dusseault2}, aside from the obvious difference that, on graphene, low coverage phases are ordered. \\ \indent 
The remarkable quantitative agreement between our results and the experimental observations of Ref. \cite{ikr} might seem surprising, given the crudeness of the silica-\ph2 potential adopted here; however, it is merely a consequence of the fact that the physics of \ph2 layering on a substrate is mostly dominated by the interactions among \ph2 molecules. The single, most important aspect of the \ph2-substrate interaction, which affects first layer promotion coverage, is its well depth $D$; the value chosen here, namely 232 K, appears to be in the right ballpark\footnote{This value is approximately 40\% of that on a graphene substrate, for which second layer promotion is predicted to occur at $\theta\sim 0.112$ \AA$^{-2}$ \cite{dusseault2}.}, given the quantitative agreement between the first and second layer coverages predicted here and those reported in Ref. \cite{ikr}. This suggest that our calculation should capture at least the main physical aspects of the system, hence assess the plausibility of the conclusions of Ref. \cite{ikr}.
\\ \indent
Next, we proceed to illustrate the results of our calculation in detail, focusing on structural and possible SF properties of the films. We begin by making the following general observation: there is {\em nothing}  fundamentally different in the behavior of adsorbed \ph2 films on this substrate, compared to other substrates investigated in the past, of widely varying strength \cite{melting,nho1,dusseault2,nho2}. The reason for this uniformity, is that the relatively large radius of the hard core repulsion of the intermolecular potential acts to suppress dramatically quantum-mechanical exchanges of \ph2 molecules. The almost complete absence of exchanges\footnote{More quantitatively, we estimate that the fraction of \ph2 molecules involved in exchange cycles in the most favorable conditions explored here, namely for the top layer of a two-layer film of coverage $\theta=0.155$ \AA$^{-2}$ at a temperature $T=0,5$ K is less than 0.001\%.}, in turn, imparts to the system a largely classical behavior, one that zero-point motion does not substantially alter. Lack of particle exchanges, besides the absence of superfluidity (known to be underlain by exchanges \cite{feynman}), causes the crystallization of the system at low temperature\footnote{It is worth pointing out that even $^4$He would crystallize at low temperature, if exchanges were suppressed, as shown in Ref. \cite{role}.}. It is worth noting that the continuous-space worm algorithm is especially efficient at sampling permutations, and allows for their observations in system in which they are relatively infrequent, such as solid helium \cite{lut}. It seems therefore reasonable to conclude that the observed lack of exchanges is not the result of algorithmic or sampling inefficiency, but rather reflects a genuine physical effect.
\\ \indent
Indeed, both a monolayer (whose behavior is virtually identical to that reported in Ref. \cite{melting} for a lithium substrate), as well as both bottom and top layers of a two-layer film, are found to be crystalline at temperatures below 6 K. Qualitatively, this can be assessed by a simple visual inspection of the many-particle configurations generated by the simulation; an example is shown in Fig. \ref{f2}, which represents a density snapshot (obtained from particle world lines) of the top layer, whose density is 0.067 \AA$^{-2}$ (the total coverage is 0.155 \AA$^{-2}$), at temperature $T=6$ K.
Although there are defects\footnote{Defects are attributable both to the relatively high temperature, as well as the difficulty of fitting two triangular lattices of different density and number of particles, i.e., the two different layers, in the same simulation cell.}, the arrangement of molecules on a triangular lattice is clear. A more precise assessment of the presence of order is provided by the reduced pair correlation function (integrated along the direction perpendicular to the substrate, see for instance Ref. \cite{toigo}) shown in Fig. \ref{f3}. Crystalline order is signaled by oscillations which persist at the highest distance accessible in the simulation, which comprises 340 particles altogether. The inset of Fig. \ref{f3} shows the same quantity, but computed at two temperatures, namely $T=0.5$ K and $T=6$ K. The fact that the results at the two temperatures are essentially superimposed, within the statistical uncertainties of this calculation, is an indication of the fact hardly anything happens to the system structurally, as the temperature is lowered from 6 to 0.5 K. The film is in a crystalline phase at the higher temperature, and remains in it all the way to $T=0$. Altogether, therefore, the results obtained in this work do not support the contention  made in Ref. \cite{ikr}  that the top layer remains in a liquid phase down to $T\sim 1$ K, and crystallizes below that temperature.
\\ \indent
Virtually unchanged in the same temperature range is also the kinetic energy of \ph2 molecules in the top layer, equal to 46.6(2) K at $T=6$ K, and 46.4 (3) K at $T=0.5$ K, again suggesting that below 6 K the system is essentially in its (crystalline) ground state, and very little happens as the temperature is lowered. This is remarkably similar to what is observed in bulk solid \ph2, at temperatures below crystallization \cite{cabrillo}. 
\section{Conclusions}\label{concl}
We have carried out extensive Quantum Monte Carlo simulations of a thin film of parahydrogen absorbed on a silica substrate. This work was motivated by recent measurements of elastic anomalies  \cite{ikr}, which were in turn interpreted as suggesting that the top layer of a thin \ph2 film might be in a fluid phase down to temperatures as low as 1 K, i.e., possibly very close to transitioning into its long sought, putative superfluid phase. 
We have made use of the most realistic model of the system presently available, and obtained estimates for first and second layer completion is quantitative agreement with those reported in Ref. \cite{ikr}. 
\\ \indent
The results of our investigation do not provide any support for the contention made in Ref. \cite{ikr} of a top layer that is liquid (or liquid-like) down to temperatures as low as $T\sim$ 1 K, and crystallizes at lower temperature. Rather, the physics observed on this substrate is both qualitatively, and at least semi-quantitatively identical with that on other substrates, of very different attractive strength, studied by simulation over the past two decades. These simulations have shaped our current theoretical understanding of the basic physics of \ph2 adsorbed films, among other things showing that {\em only} the crystalline phase exists at low temperature; that third layer promotion and evaporation occur at least concomitantly with, if not before 2D melting of the top layer; and that the role of the substrate is nearly insignificant, as the behavior of this system is mainly influenced by the interaction among \ph2 molecules. None of this is really contested, or controversial.
\\ \indent
Obviously, all of these considerations {\em a fortiori} apply to heavier isotopes of \ph2, i.e., HD and D$_2$; in particular, the virtual absence of quantum-mechanical exchanges in this system renders the observation of effects of quantum statistics in this system, and in general in bulk molecular hydrogen, an essentially unviable proposition. Nanoscale size clusters remain the most plausible playground for that kind of investigation.
This is, of course, not to say that the observed elastic anomalies reported in Ref. \cite{ikr}, about which the  study presented here affords no obvious insight, are not interesting and worthy of further investigation. However, the connection drawn therein between those anomalies and the possible liquid-like behavior of the top \ph2 layer, and its proximity to a SF transition, seems to lack at this time an adequate theoretical foundation. Indeed, the fact that similar elastic anomalies have been observed \cite{ikr} also in adsorbed films of Ne, not regarded as a plausible candidate to display liquid-like behavior at low temperature, much less superfluidity, suggests that they may be entirely  unrelated to such a phenomenon.

\section*{Acknowledgments}
This work was supported by the Natural Sciences and Engineering Research Council of Canada.

\section*{Conflict of interest}
The author declares that he has no conflict of interest.

\end{document}